\documentstyle[aps,epsf,epsfig,preprint]{revtex}
\def\prl#1#2#3{{ Phys. Rev. Lett.} {\bf #1}, #2 (#3)}
\def\ibid#1#2#3{{\it ibid.}  {\bf #1}, #2 (#3)}
\def\pla#1#2#3{Phys. Lett. A {\bf #1}, #2 (#3)}
\def\pre#1#2#3{Phys. Rev. E {\bf #1}, #2 (#3)}
\def\pra#1#2#3{Phys. Rev. A {\bf #1}, #2 (#3)}

\def\jsp#1#2#3{J. Stat. Phys.  {\bf #1}, #2 (#3)}

\def\physd#1#2#3{Physica D {\bf #1}, #2 (#3)}

\def\etc{ etc.}

\def\etl{$et ~al.$~}
\def\MLE{$\Lambda$}
\def\epsilonp{\epsilon^{\prime}}

\def\beqr{\begin{eqnarray}}
\def\eqnr{\end{eqnarray}}
\def\beq{\begin{equation}}
\def\bc{\begin{center}}
\def\ec{\end{center}}
\def\eqn{\end{equation}}
\begin{document}
\title{Bifurcations and transitions in the quasiperiodically 
driven logistic map}
\author{Surendra Singh Negi, Awadhesh Prasad and Ramakrishna Ramaswamy} 
\address{School of Physical Sciences\\ Jawaharlal Nehru
University, New Delhi 110 067, INDIA}
\date{\today}
\maketitle
\begin{abstract}
We discuss several bifurcation phenomena that occur in the 
quasiperiodically driven logistic map. This system can
have strange nonchaotic attractors (SNAs) in addition to 
chaotic and regular attractors; on SNAs the dynamics is aperiodic,
but the largest Lyapunov exponent is nonpositive. There are a
number of different transitions that occur here, from periodic
attractors to SNAs, from SNAs to chaotic attractors, etc. We describe
some of these transitions by examining the behavior of the largest Lyapunov
exponent, distributions of finite time Lyapunov exponents and
the invariant densities in the phase space.
\end{abstract}

\newpage
\section{INTRODUCTION}

Quasiperiodically driven mappings where the dynamics can be strange
(namely on fractal attractors) but nonchaotic (namely having a lack
of sensitivity to initial conditions) are of considerable current
interest \cite{rmp,general}.  Strange nonchaotic attractors (SNAs)
were first described by Grebogi \etl in 1984 \cite{gopy}, and are now
known to be generic in quasiperiodically driven systems. Such
dynamics is paradoxical in some ways. The motion on SNAs is
aperiodic, but over long times, nearby trajectories will coincide.
The dynamics is almost always characterized by intermittency, which
is indicative of the fact that such attractors are highly nonuniform
and have a complicated interweaving of (locally) stable and unstable
regions.

External forcing allows for a additional means of probing nonlinear
dynamical systems.  If the forcing is periodic, then the motion of
the system becomes either periodic or chaotic, but for quasiperiodic
forcing (for example when a system is modulated with two frequencies
which are incommensurate with each other) SNAs become possible.
Strange nonchaotic dynamics usually occurs in the vicinity of strange
chaotic behavior and periodic or quasiperiodic (nonstrange,
nonchaotic) behavior. As a result, such systems can show transitions
between dynamical states and bifurcation phenomena which are similar
to those in analogous autonomous systems (for example the several
scenarios \cite{eckmann81} such as the period-doubling route to
chaos, intermittency, attractor crises, \etc) as those that are
distinct from the bifurcations of unforced systems
\cite{prl,pre,ott}.  These have been extensively studied in a
number of different contexts: questions of interest range from the
mechanisms through which SNAs are created
\cite{grebogi87,rboag,kpf,hh,k,frac,yl,blowout,lai}, how they
may be characterized \cite{pf,ditto},
experimental systems where these might occur
\cite{ditto,bulsara,newexp}, etc.

In this paper we examine the transitions between a number of
different types of attractors in the quasiperiodically driven
logistic map. In this prototypical driven system,
the attractors can be strange and nonchaotic, in addition to being 
strange and chaotic or nonchaotic and regular (torus attractors). 
Studies of the driven logistic map have played an important
role in the study of SNAs. Kaneko \cite{k} first observed ``torus
wrinkling'' in this system: this was eventually described
as the fractalization route to SNAs \cite{frac}. The creation of 
SNAs through the collision of stable period--doubled tori with 
their unstable parent torus was also first studied in this 
system \cite{hh}. One great advantage in studying the driven
logistic mapping (even though it is somewhat difficult
to obtain analytic results) is that the 
undriven logistic map has been extensively studied over the past
few decades. Many of the features of the driven system find
their parallels in the undriven system. At the same time, however,
since the
dynamics in the logistic map is generic of a wide class,
the behaviour that can be simply studied in the driven logistic
map is characteristic of most driven nonlinear systems. 
Some scenarios through which SNAs are formed in this and related systems
have been reviewed recently \cite{rmp}.

SNAs occur in several different parameter ranges, between
regions of periodic or torus attractors and regions of chaotic
attractors. There is a plethora of possible dynamical 
transitions, some of which have parallels in the undriven
system, such
as torus bifurcations, $$\ldots n~\hbox{T} \leftrightarrow
2n~\hbox{T}\leftrightarrow
\ldots, $$ and others which do not, such as
transitions from tori to SNAs
$$\ldots n~\hbox{T} \leftrightarrow n~\hbox{ band SNAs}\leftrightarrow 
\ldots $$
$$\ldots 2^n~\hbox{T} \leftrightarrow 2^{n-1}~\hbox{ band
SNAs}\leftrightarrow \ldots .$$ SNAs merge in a manner similar to
the case of reverse bifurcations, SNAs widen, in a manner similar
to widening chaotic crises,  and can 
transform from one type to another or from SNAs to
chaotic attractors.

The main focus of this paper, in addition to characterizing 
the above bifurcations and transitions in this
system through the Lyapunov exponent and its fluctuations, is also to
examine the manner in which the invariant measure varies with the
parameters of the system, and the effect that this has on the
dynamics. 
In many instances, these bifurcations or transitions involve the
Lyapunov exponent going through zero, and can be analyzed in terms of
symmetries in the tangent--space dynamics. 

In Section II, we
introduce the model, and discuss the analysis in terms of global and
local Lyapunov exponents, and the return map for stretch exponents.
The the characteristic behavior of the Lyapunov exponents 
at the various transitions to different attractors are discussed
in Sec.~III. This is followed by a summary in Sec.~IV.

\section{Model}

We study the logistic map with quasiperiodic modulation of the
parameter $\alpha$,
\beqr
\label{logistic}
     x_{n+1} &=& \alpha [1 + \epsilon \cos 2\pi
     \theta_n] x_n (1 - x_n) \\ 
\theta_{n+1}  &=&  \theta_n + \omega \mbox{~mod~} 1, 
\eqnr
where $\omega$ is taken to be an irrational number (usually the
golden mean ratio, $ (\sqrt{5}-1)/2$. Successive iterates of
$\theta$ will densely and uniformly cover the unit interval in a
quasiperiodic manner, and the system therefore has {\it no} periodic
orbits.  As in our previous studies \cite{prl,pre}, we rescale the parameter
$$\epsilonp =\frac{\epsilon}{4/\alpha-1}$$ for convenience, and
study the system in the range $2 \le \alpha \le 4$ and $0 \le 
\epsilonp~\le~1$.

The Lyapunov exponent corresponding to the $\theta$-rotation
is trivially zero; however the other exponent, which is of
most importance in determining the dynamics varies with the
parameters. It can be calculated by averaging the {\it stretch exponents}
tangential to the flow, namely 
\beq
\label{stretch}
y_n = \ln \vert \alpha [1 + \epsilon \cos 2\pi
     \theta_n] (1 - 2 x_n) \vert ,
\eqn
which is the local derivative of the mapping
along a trajectory.  The local or $N$--step Lyapunov exponent is
\beq
\label{nlyap}
\lambda_N  =  \frac{1}{N} \sum_{j=1}^N y_j
\eqn
from which, asymptotically, one gets the global Lyapunov exponent
\beq
\Lambda = \lim_{N\to \infty} \lambda_N.
\label{lyap}
\eqn

The stretch exponents are negative in the region bounded by the curves 
\beq
\label{xtheta}
 x_{\pm}(\theta) = {1\over 2} \left [1\pm {1\over{
\alpha(1+\epsilon \cos 2\pi\theta)}}\right ].
\eqn
Any change in the dynamics such that the invariant measure 
is increased in the region $[x_-(\theta), x_+(\theta)]$
will therefore lead to a decrease in the
Lyapunov exponent, and, conversely, depletion of 
measure in this region will
naturally lead to an increase in the Lyapunov exponent.  

Note, however, that the invariant measure, 
$\rho_{\alpha,\epsilon}(x,\theta)$, for 
this mapping is not known exactly (except at $\alpha=4, \epsilon = 0$).
It is therefore 
determined numerically by partitioning the phase space $(x,\theta)$
into bins and examining the itinerary of a long trajectory.
Shown in Fig.~1(a) is an example of a SNA; the corresponding invariant
measure is shown in Fig.~1(b).  The solid line is the locus of 
$x_{\pm}(\theta)$, (cf. Eqn.~(\ref{xtheta})), showing that the
SNA is largely located within the contracting regions in phase space, but
also has considerable support in the unstable regions.

Local LEs, $\lambda_N$, depend on initial conditions but (with
probability 1) $\Lambda$ does not \cite{multi}. In order to characterize
the non--uniformity of the attractor, it has proved instructive to 
examine the distribution of local Lyapunov exponents. The probability
density,
\beqr P(N,\lambda) d\lambda &=& {\rm probability~ }~{\rm that~} \lambda_N
{\rm ~lies~ between}~ \nonumber \\
     & &\lambda ~ {\rm and}~ \lambda + d\lambda,
\eqnr
has been seen to have characteristic limiting forms that depend
on the nature of the attractor \cite{pr,trichy}. Of course, as $N
\to \infty$, $P(N,\lambda) \to \delta(\lambda-\Lambda)$, but
the nature of the finite--size corrections and the approach to the
limit are distinctive for different dynamical states.  This analysis
is relevant for the study of nonuniform attractors, particularly for
different attractors across crisis points or at intermittency,
when stretched exponentials often occur \cite{pr}.

\section{Results and Discussion}

In this Section we study the variation of the dynamics through
several transitions in the system as the parameters $\alpha$ and
$\epsilonp$ are varied. Phase diagrams for this system in different
parameter ranges have been obtained in a number of previous studies 
\cite{prl,pre,kpf,hh,trichy,witt}, and it is known that there are two distinct
regions of chaotic dynamics, corresponding, respectively to high driving
(large $\epsilonp$) and large nonlinearity (large $\alpha$). These are
shown in Fig.~2 as C$_2$ and C$_1$ \cite{prl}, and they 
separate a region of quasiperiodic (torus) dynamics. Strange nonchaotic motion 
occurs on the boundaries of these regions \cite{pre} as shown in Fig.~2.

In addition to the asymptotic nontrivial Lyapunov
exponent \MLE, we also examine the distribution of 
$N$--step Lyapunov exponents, and the variance
of this distribution. 

The variation in \MLE~with $\alpha$~for fixed $\epsilonp$
is shown in Fig.~3(a). Although there are several
bifurcations and transitions, these are not easily visible in
the behavior of the Lyapunov exponent. We therefore examine
an approximate or partial bifurcation 
diagram which can be obtained for this system by plotting the values 
of $x$ that obtain within a narrow window in $\theta$, namely
in the interval ($\theta, \theta + d\theta$). This
will depend upon the choice of $d\theta$ and also on
the particular value of $\theta$ chosen, but qualitative
features of the bifurcation diagram are not affected. 

The partial bifurcation diagram, Fig.~3(b), shows some of the
transitions clearly:
the torus--doubling, for instance, and the transition from
torus attractors to fractal attractors. Since singularities
are dense in $\theta$, a SNA or a chaotic attractor shows up
as a spread of points (see Fig.~3(b), for example), while
a torus attractor appears as a point or a finite set of points.
The distinction between 
a chaotic attractor or a SNA is not evident in the bifurcation
diagram, but examining Figs.~3(a) and (b), and the change in
the variance in \MLE, shown in Fig.~3(c) gives a complete
picture of the different attractors that are present in the system.
(Our calculations of the variance, $\sigma$, are from
$50$ samples, each of total length $10^6$ iterations.)

Quasiperiodic forcing converts the fixed points of the logistic map
into tori.  At the period--doubling bifurcation, the Lyapunov exponent 
is exactly zero, as in the unforced system, but 
with quasiperiodicity, the sequence of torus--doublings is interrupted
and leads to SNAs, either through the collision of stable and unstable
tori as discussed by Heagy and Hammel \cite{hh}, or through
fractalization \cite{frac,kk}. This latter scenario is the most
common route to SNA, and since there is,
apparently, no bifurcation involved, it is not clearly understood
as to how a torus gets increasingly wrinkled and transforms into 
a fractal attractor in the process. Other routes to SNA are 
known, some of which, like intermittency \cite{prl,venkat}, occur in this 
system and others, such as the blowout bifurcation route
\cite{yl}, which do not. 
    
\subsection{From SNAs to SNAs: Merging and widening crises}
\label{ss}

As has been remarked earlier, several 
 of the bifurcation phenomena of (unforced) chaotic dynamical
systems find their parallels in quasiperiodically forced systems.
For example, there are analogue of crisis phenomena. In the
present system SNAs have a certain number of
``bands''.  $n$--band SNAs are formed from $n$-tori via
fractalization, or from 2$n$--tori through the Heagy--Hammel
mechanism.  Additionally, one--band SNAs can 
form from a 1--torus \cite{prl} or a 3--torus \cite{trichy,witt}
due to  saddle-node bifurcations.  

As parameters are further varied, SNAs themselves evolve and merge at
quasiperiodic analogues of band--merging crises or reverse
bifurcations: $n$--band SNAs transform to $n/2$--band SNAs.  Through
such a transition, when the dynamics remains nonchaotic and strange,
\MLE~ is a good order parameter.  Sosnovtseva \etl~\cite{sosno}, who
discovered an example of this transition in the driven H\'enon and
circle maps, demonstrated the merging by examining the phase
portrait.  Given the fairly narrow range over which SNAs exist in any
system, such transitions also occurs in a restricted range, and often
such crises can occur after the transition to chaotic attractors.
However, the variation of \MLE~does not follow a uniform pattern as
in the unforced case \cite{crutchfield,mr}. 

SNAs which are formed via fractalization may merge at negative \MLE.  The
Lyapunov exponent {\it {decreases}} with increasing nonlinearity;
shown in Fig.~4(a) is the variation of \MLE~at a band--merging
bifurcation (shown by the arrow) from a two--band SNA to a one--band
SNA for $\epsilon=0.05$. The critical value of the parameter is
$\alpha_c\approx 3.387 439$ and the corresponding `partial' 
bifurcation diagram of each alternate iterate of $x$
is shown in Fig.~4(b). At merging crises in unforced systems,  
the Lyapunov exponent usually does not show an increase (unlike at
widening crises) while with forcing, the exponent actually decreases.

This can be understood by examining the invariant density on the
attractor before and after merging. The two bands of the two--band SNA
straddle the region of phase space bounded by the curves $x_{\pm}(\theta)$,
Eq.~\ref{xtheta}.
After the merging transition, the density is enhanced mainly in this
region where the stretch exponents are all negative. As
a consequence, the Lyapunov exponent {\em must} decrease. 
(A plot of the difference in the invariant
density on the SNAs before and after merging is shown in Fig.~4(c).)
This appears to be the typical behaviour at the SNA
band--merging transition \cite{private}, and is in contrast
to the behaviour of the unforced system.

Widening crises also occur, where the SNA abruptly changes
size as a parameter is varied (near W$_1$ in Fig.~2). Shown in Fig.~5 is such an
example, which occurs on the line $\epsilonp=1$. Here, 
the post--crisis SNA, shown in Fig.~5(b), is clearly larger
than the pre--crisis SNA (Fig.~5(a)) which is created via 
fractalization. The signatures of the transition are evident 
in both \MLE~and 
the variance [Figs.~5(c-d)]. The sudden expansion of the attractor
seems to be due to collision with the unstable saddle (which also
gives rise to the intermittency transition to SNA \cite{prl}).
This saddle is difficult to locate, but in the three--dimensional
extension of this system, namely the quasiperiodically driven
H\'enon map, Osinga and Feudel have recently described similar
crises in detail \cite{osinga} and have obtained the saddle there
as well. In contrast to merging crises, the density is now 
enhanced in the unstable regions of phase space: \MLE~therefore increases. 
The distribution of finite---time LEs also has a stretched 
exponential tail which is characteristic of the
crisis--induced intermittency \cite{pr}
[see Fig.~5(e)]. Similar interior crises also take place
for lower forcing, when a fractalized SNA transforms to an
intermittent SNA. An example of this is shown in Fig.~6,  
near $\epsilonp=0.614...$ and $\alpha_c=3.13188...$ (W$_2$ in Fig.~2),
namely on the edge of region C$_2$. 

\subsection{From SNAs to Chaos}

On variation of a parameter, a SNA can be transformed into
a chaotic attractor if the largest Lyapunov exponent becomes
positive.  This transition, which is also not associated with any
bifurcation, is not as dramatic as from the torus to a SNA. Typically
there is a smooth variation in the Lyapunov exponent.

At the SNA$ \to $ chaos transition, all the Lyapunov exponents 
of the system are zero. Since \MLE~ [Eq.~(\ref{lyap})] is a 
global average, it is of interest to know how the individual terms in
Eq.~(\ref{stretch})  cancel out so as to give a value zero.
Consider the partial finite sums \cite{insa},
\beq
\label{sum1}
\lambda_N^+ =  {1 \over N_+} \sum_{i} y_i ,~~~~~~ y_i > 0,  
\eqn
and
\beq
\label{sum2}
\lambda_N^- =  {1 \over N_-} \sum_{i} y_i ,~~~~~~ y_i < 0,  
\eqn
namely the separate contributions to the local Lyapunov exponent.
These are obtained by partitioning a trajectory 
into $N_+$ points on expanding regions, where the 
stretch exponents are positive,  and $N_-$ points on 
contracting regions where the stretch exponents are negative with
$N=N_++N_-$. Clearly $\lambda_N = \frac{N_+}{N}\lambda^++\frac{N_-}{N} 
\lambda^-$, and with the limits
\beq
\lim_{N\to\infty} \lambda_N^{\pm} \to \lambda^{\pm}
\eqn
(cf. Eqs.~(\ref{nlyap}) and (\ref{lyap})) $\Lambda = \lambda^++\lambda^-$.

If there are symmetries in the system \cite{prss,newsymm},
then $\lambda^+$ and $\lambda^-$ can equal each other and thereby
yield $\Lambda = 0$. In such situations, there is a term--by--term
cancelation of positive and negative stretch exponents \cite{newsymm}. 
At torus bifurcations, for instance, or at the blowout bifurcation
transition to SNAs \cite{gopy,yl} this situation applies.

On the other hand, there can be an ``accidental''
equality, namely, since both $\lambda^+$ and $\lambda^-$ will
be functions of parameters $\alpha$ and $\epsilon$, it
may happen that they are both equal in magnitude for some
choice of parameters, and thereby lead to a zero value for the
Lyapunov exponent.

At the SNA to chaos transition, this latter situation obtains.
Lai \cite{lai} investigated the transition from SNAs to chaotic
attractors and found that this transition occurs only when the
contraction and expansion rates for infinitesimal vectors along a typical
trajectory on the attractor (namely  $\lambda^+$ and $\lambda^-$)
become equal; the LE passes through zero {\it linearly} as in Fig.~7(a).

There are, however, subtler effects that become apparent when
examining fluctuations in the local Lyapunov exponents. The variance
of the distribution shows a small increase across the
transition [Fig.~7(b)]. 
There is an increase in the fluctuations in the transition from
torus attractors to SNA as well (from essentially zero to some finite
value). At the SNA to chaos transition, the scale of fluctuations
doubles, and is therefore noticeable on this scale.
The SNA $\to$ Chaotic attractor transition is
not accompanied by any major change in the form or shape of the
attractor, and the Lyapunov exponent itself changes only from being
negative to positive. Yet the fluctuations on the chaotic attractor 
are always larger than those on the nonchaotic attractor.

This enhancement in fluctuations in \MLE~ can be analyzed via the
invariant density on the attractors. Shown in Fig.~8 is the 
{\it difference} in the invariant density on a SNA and a chaotic attractor
symmetrically placed about the SNA $\to$ Chaos transition. While it
is clear that the morphology of the attractor does not change
significantly, it can also be seen that on the SNA,
the invariant density is enhanced in the regions where stretch exponents 
are mainly negative. Since
these span the range from $-\infty$ to 0 within the region
$[x_-(\theta), x_+(\theta)]$, they contribute
significantly to the variance without greatly affecting the mean:
the Lyapunov exponent, thus, does not change significantly, but
the fluctuations show an increase.

Crises which occur {\em after} the transition to chaos appear to be
very similar to analogous phenomena in systems without forcing
\cite{grebogi87,mr};
$\Lambda$ has a power--law dependence on the parameter,
$\Lambda - \Lambda_c \propto  (\alpha-\alpha_c)^\beta$. For example
the  widening crisis due to saddle node bifurcation
along $\epsilon^{\prime}=0.02$, across the
transition point $\alpha_c=3.85226..$ is observed where the value of  $\beta$
 is $\approx 0.67$ which is larger than the unforced case \cite{mr}.
\section{Summary}

In this paper we have studied different dynamical transitions
that occur in the quasiperiodically driven logistic map, the main
focus being on the quasiperiodic analogues of crisis phenomena. 

SNAs are formed via several different routes, and coexist with 
chaotic as well as other nonchaotic attractors.  As a result there
are interesting transformations of one type of attractor to another.
The torus to SNA transitions have been extensively described 
previously \cite{rmp,pre}.
Fractalization, which is the most common scenario for the formation 
of SNAs is a ``P2C2E'' \cite{p2c2e}: no explicit bifurcation mechanism
has been identified with this route. However, fractalized SNAs can
be transformed into intermittent SNAs upon collision with an unstable
torus in an analogue of the interior crisis, and also undergo merging
at the analogue of a merging crisis. 

In all these transitions, the
nontrivial Lyapunov exponent is a good order--parameter, and 
furthermore, its fluctuations the distributions of finite--time
exponents provide additional signatures for these transitions.
In order to understand the nature of the variations of these
quantities, however, it is necessary to study the
the invariant measure on SNAs. We show that they have
support even in regions which are locally unstable. 

At many of the
transitions that involve such attractors, there is no particular change
in form or morphology. However, by examining differences in the invariant
density at different parameter values, it becomes clear that at
these transitions, the manner in which the dynamics explores
the attractor can change drastically, and small
changes in the density can lead to significant effects in the 
fluctuation properties of such attractors.

Typical attractors in dynamical systems are nonuniform, both with 
respect to the natural invariant measure as well as in the rate 
at which nearby trajectories---locally---diverge or converge. 
If the measure on converging regions exceeds that on the diverging 
regions, then the attractor is asymptotically nonchaotic and will 
be characterized by a negative Lyapunov exponent. External 
modulation can be one method of
altering the invariant measure, and this method of 
creating nonchaotic dynamics \cite{insa,wong} therefore provides a new
means of synchronization and control. In this context,
therefore, studies of the bifurcations and transformations of 
SNAs give further insights into the interplay between global stability
and local instability.

\centerline{\bf ACKNOWLEDGMENT} This research was supported
from the Department of Science and Technology, India.

\newpage
\section*{Figure Captions}

\vskip0.5cm

\noindent {\bf Figure 1:} (a) A SNA in the driven logistic map
at $\epsilon^{\prime}=0.75$. and $\alpha= 3.26774$, and
(b) the corresponding invariant density (scaled $\times 10^{3}$) in
the phase space. The solid lines (in (a)) and the dashed
lines (in (b)) are the loci of
$x_{\pm}(\theta)$, Eqn.~(\ref{xtheta}). \\

\ \\
{\bf Figure 2:}Schematic phase diagram for a small region in parameter 
space, for the forced logistic map see details in Ref~.\cite{pre}
. T and C correspond to regions of torus and chaotic attractors.  
SNAs are mainly found in the shaded region along the boundary of 
T and C (marked S). W denotes the region where widening crises 
occur (see the txt). \\

\ \\
{\bf Figure 3:} (a) Variation of \MLE~ as a function of $\alpha$ for fixed
$\epsilon^{\prime} = 0.595$. Note the highly oscillatory structure
indicative of several transitions in the system. These are clearly
shown in (b) which is a `partial' bifurcation diagram  with
$\theta=0.5$ and $d\theta=0.0001$. The regions of torus attractors
as well as strange behavior can be easily seen.
(c) The variance in \MLE.  The symbol T and C correspond to torus and
chaotic attractors while S stands for SNA.  \\

 \ \\
{\bf Figure 4:}Variation of \MLE~at a band--merging bifurcation (shown by the arrow)
from a two--band (fractalized) SNA to a one--band 
(Heagy--Hammel) SNA along $\epsilon=0.05$ at $\alpha_c
\approx 3.387 439$.  (b) The corresponding `partial' bifurcation diagram
of the second iteration for $x$ in the interval
($\theta,\theta+d\theta$), $\theta=0.3$ and $d\theta=0.01$, and 
(c) the difference in densities (scaled $\times 10^3$) after merging 
($\alpha=3.389$) and SNA before merging ($\alpha=3.386$) 
along $\epsilon=0.05$. The subscripts 2bs and 1bs refer, respectively,
to 2--band and 1--band SNAs. \\
 \ \\
{\bf Figure 5:} (a) The SNA before widening along $\epsilon^{\prime}=1$ at
$\alpha=3.4449$ and (b) the SNA after the widening crisis at $\alpha=3.44498$;
(c) the variation of Lyapunov exponent, $\Lambda,$ across the transition 
point $\alpha_c=3.444955..$; (d) the variance
in \MLE, and (e) the distribution of finite-time Lyapunov exponents at
$\alpha=3.44498$, showing the stretched exponential tail which is
characteristic of intermittency.  \\
\ \\
{\bf Figure 6:}(a) The fractalized SNA (before widening) along 
$\epsilon^{\prime}= 0.614$ at $\alpha=3.13192$ and (b) the intermittent 
SNA after the widening crisis at $\alpha=3.13184$.  \\
\ \\
{\bf Figure 7:}The transition from SNA to a chaotic attractor along
$\epsilon^{\prime}=0.3$. (a) The Lyapunov exponent across the
transition, and (b) its fluctuations. \\
\ \\
{\bf Figure 8:}The difference in densities (scaled $\times 10^3$)
between a chaotic attractor ($\alpha=3.515$) and SNA
($\alpha=3.508$) along $\epsilon^{\prime}=0.3$.
The plot of $x_{\pm}(\theta)$ (for $\alpha=3.508$)
of Eqn.~\ref{xtheta} is superimposed as a
dashed line.\\


\begin{thebibliography}{99}
\bibitem{rmp}  A. Prasad and R. Ramaswamy, to be published.
\bibitem{general} J. P. Eckmann and D. Ruelle, Rev. Mod. Phys. {\bf 57}, 
617 (1985).
\bibitem{gopy} C. Grebogi, E. Ott, S. Pelikan, and J. A. Yorke,
\physd{13}{261}{1984}.
\bibitem{eckmann81} J. P. Eckmann, Rev. Mod. Phys. {\bf53}, 643 (1981).
\bibitem{prl} A. Prasad, V. Mehra, and R. Ramaswamy, \prl{79}{4127}{1997}.
\bibitem{pre} A. Prasad, V. Mehra, and R. Ramaswamy, \pre{57}{1576}{1998}.
\bibitem{ott} E. Ott, {\it Chaos in dynamical systems}, (Cambridge
University Press, Cambridge, England, 1994).
\bibitem{grebogi87} C. Grebogi, E. Ott, F. J. Romeiras, and J. A. Yorke, \pra{36}{5365}{1987}.
\bibitem{rboag} F. J. Romeiras, A. Bondeson, E. Ott, T. M. Antonsen,
and C. Grebogi, \physd{26}{277}{1987}.
\bibitem{kpf} S. P. Kuznetsov, A. S. Pikovsky, and U. Feudel, \pre{51}{R1629}{1995}. 
\bibitem{hh} J. F. Heagy and S. M. Hammel, \physd{70}{140}{1994}.
\bibitem{k} K. Kaneko, Prog. Theor. Phys., {\bf 71}, 1112 (1984).
\bibitem{frac} T. Nishikawa and K. Kaneko, \pre{54}{6114}{1996}.
\bibitem{yl} T. Yal\c{c}inkaya  and Y. C. Lai, \prl{77}{5039}{1996}.
\bibitem{blowout} E. Ott and J. C. Sommerer, \pla{188}{39}{1994}; P.
Ashwin, J. Buescu, and I. Stewart, \ibid{193}{126}{1994}; Y. C. Lai
and C. Grebogi, \pre{52}{R3313}{1995}. 
\bibitem{lai} Y. C. Lai, \pre{53}{57}{1996}; Y. C. Lai, U. Feudel, and C.
Grebogi, \ibid{54}{6070}{1996}.
\bibitem{pf} A. Pikovsky and U. Feudel, Chaos, {\bf 5}, 253 (1995).
\bibitem{ditto} W. L. Ditto, M. L. Spano, H. T. Savage, S. N. Rauseo,
J. Heagy, and E. Ott, \prl{65}{533}{1990}.
\bibitem{bulsara} T. Zhou, F. Moss, and A. Bulsara, \pra{45}{5394}{1992}.
\bibitem{newexp} W. X. Ding, H. Deutsch, A. Dinklage, and C. Wilke,
\pre{55}{3769}{1997}; Z. Liu and Z. Zhu, Int. J. Bifurcation and Chaos, {\bf 6}
1383 (1996); Z. Zhu and Z. Liu, \ibid{7}{227}{1997}; A. J. Mandell and K. A.
 Selz, \jsp{70}{255}{1993}.
\bibitem{multi} Except for systems which have multiple coexisting
attractors, such as the double kicked--rotor, Duffing oscillator, \etc
\bibitem{pr} A. Prasad and R.  Ramaswamy, \pre{60}{2761}{1999}
and references therein.
\bibitem{trichy} A. Prasad and  R. Ramaswamy,  {\it Finite-time Lyapunov
Exponents of Strange Nonchaotic Attractors}, 
in {\it  Nonlinear Dynamics: Integrability and Chaos}
Eds. M Daniel, K Tamizhmani and R Sahadevan (Narosa, New Delhi,
2000), pp. 227--34.
\bibitem{witt}  A. Witt U. Feudel, and A. S. Pikovsky, \physd{109}{180}{1997}.
\bibitem{kk}K. Kaneko, Prog. Theor. Phys. {\bf 72}, 202 (1984).
\bibitem{venkat} A. Venkatesan, M. Lakshmanan, A.  Prasad, and R.
Ramaswamy, Phys. Rev. E, in press 2000.
\bibitem{sosno} O. Sosnovtseva, U. Feudel, J. Kurths, and A. Pikovsky,
\pla{218}{255}{1996}. 
\bibitem{crutchfield} J. Crutchfield, D. Farmer, and B. Huberman, Phys. Rep.,
{\bf 92}, 45 (1982).
\bibitem{mr} V. Mehra and R. Ramaswamy, \pre{53}{3420}{1996}. 
\bibitem{private} We have studied the forced ring map and H\'enon maps and
find that at SNA mergings in these systems, the largest Lyapunov exponent
decreases in a manner very similar to that in the logistic system.
\bibitem{osinga} H. Osinga and U. Feudel, ``Boundary
crisis in quasiperiodically forced systems'', Physica D, in press 2000.
\bibitem{insa} A. Prasad and R. Ramaswamy, `` Can strange nonchaotic 
attractors be created through stochastic driving?'', Proc. Ind. 
Nat. Sci. Acad., in press 2000.
\bibitem{prss} A. Prasad, R. Ramaswamy, I. I.  Satija and N. Shah, 
\prl{83}{4530}{1999}.
\bibitem{newsymm} A. Prasad, S. S. Negi, S. Datta and R. Ramaswamy, 
in preparation. Note that the symmetries can be hidden as in Ref.~\cite{prss}
or complicated, as at tangent bifurcations. 
\bibitem{p2c2e} A ``Process too Complicated to Explain'' from 
S.~Rushdie, {\it Haroun and the Sea of Stories}, Penguin, 1990, pp.~57.
\bibitem{wong} J. W. Shuai and K. W. Wong, \pre{59}{5338}{1999}.

\end{thebibliography}
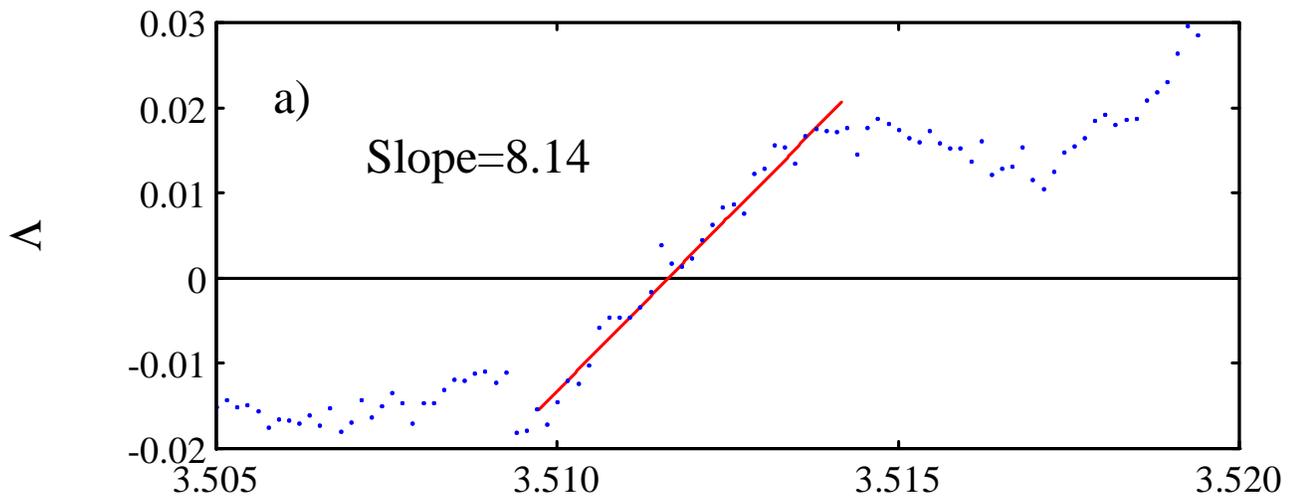
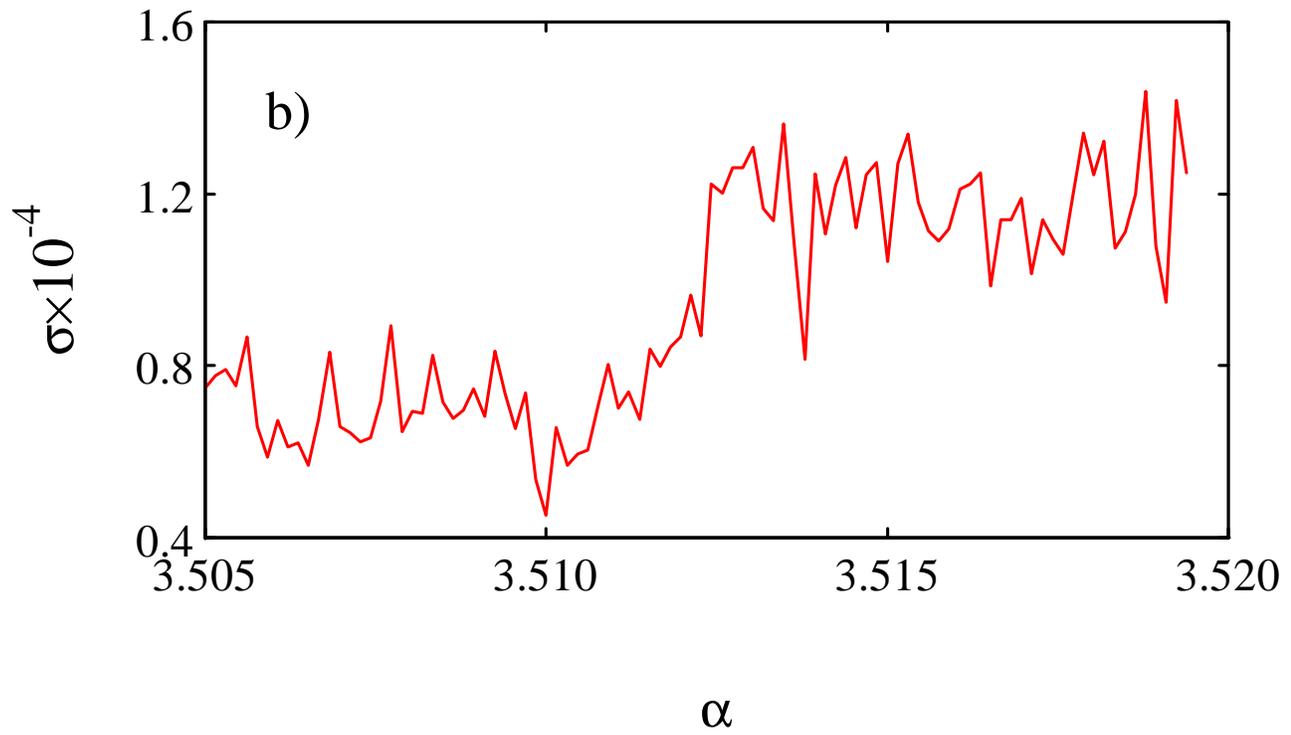
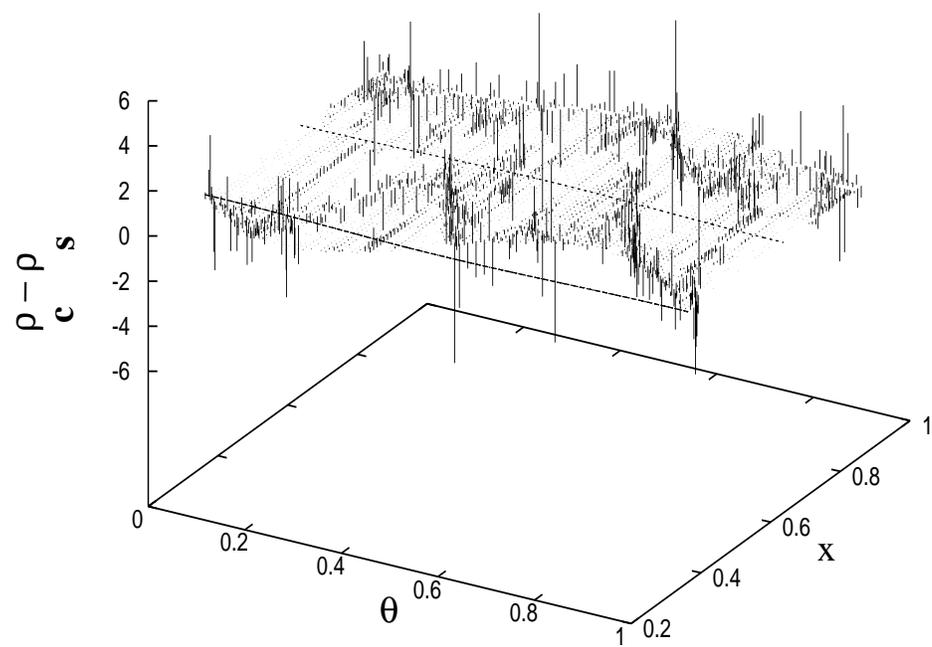
\end{document}